# Data-Driven Estimation of the interfacial Dzyaloshinskii–Moriya Interaction with Machine Learning

Davi Rodrigues[1,*], Andrea Meo[1], Ali Hasan[2],
Edoardo Piccolo[1], Adriano Di Pietro[3], Alessandro Magni[3], Marco Madami[4],
Giovanni Finocchio[2], Mario Carpentieri[1], Michaela Kuepferling[3], and Vito Puliafito[1,*]

[1] *Department of Electrical and Information Engineering, Politecnico di Bari, 70126 Bari, Italy*
[2] *Department of Mathematical and Computer Sciences, Physical Sciences and Earth Sciences, University of Messina, 98166, Messina, Italy*
[3] *Istituto Nazionale di Ricerca Metrologica, 10135, Torino, Italy*
[4] *Department of Physics and Geology, University of Perugia, 06123 Perugia, Italy*

**Abstract**

Machine learning offers powerful tools to support experimental techniques, particularly for extracting latent features from large datasets. In magnetic materials, accurately estimating the interfacial Dzyaloshinskii–Moriya interaction strength remains challenging, as existing experimental methods often rely on indirect measurements and can yield inconsistent results across techniques. Because this interaction is often extracted experimentally from bubble domain expansion, we investigate whether bubble textures alone contain sufficient and reliable information for data driven DMI inference. We therefore develop a compact convolutional neural network trained on a comprehensive micromagnetic dataset of magnetic bubble domains designed to emulate magneto optical Kerr effect imaging, including structural non uniformity, additive noise, and image pixelation. The proposed network demonstrates strong robustness against sample inhomogeneities, noise, and reduced spatial resolution. Furthermore, it exhibits reliable generalization by accurately predicting DMI values outside the trained interval. These results support the use of machine learning as a fast and quantitative tool to characterize magnetic textures with interfacial DMI.

*corresponding authors: davi.rodrigues@poliba.it, vito.puliafito@poliba.it

## INTRODUCTION

Machine learning (ML) is rapidly becoming a practical instrument in materials science [1–4]. ML enables to extract latent structures from complex, noisy, and high-dimensional data streams (images, spectra, microscopy movies) that are otherwise difficult to interpret with hand-engineered descriptors. This paradigm is particularly impactful when the relevant physical parameter is not directly observable, but, instead, it manifests indirectly through subtle, distributed signatures in experimental data [3–5].

In magnetic thin-film magnetism, a prominent example of latent physical phenomena is the Dzyaloshinskii–Moriya interaction (DMI) [6]. This interaction is an antisymmetric exchange arising from broken inversion symmetry and spin-orbit coupling. DMI establishes a preferred chirality for spatial variations in magnetization, which underpins the stabilization and dynamics of chiral magnetic textures, such as chiral domain walls, vortices, and skyrmions [7–10]. Therefore, quantifying DMI reliably is central to micromagnetic modeling and for optimizing multilayer stacks and interfaces [11–13]. However, accurately measuring DMI is challenging [8,14,15]. Widely used techniques, such as asymmetric bubble expansion in the creep regime [15], Brillouin light scattering (BLS) [16], and other domain-wall or spin-wave-based protocols [17], often yield different values for similar materials. This is due to systematic uncertainties arising from model assumptions, pinning/roughness, and the need for independently known magnetic parameters.

A recent international round robin comparison of BLS and asymmetric bubble expansion revealed that domain wall roughness, sample and interface disorder, and the selection of creep model extension can significantly impact the DMI values obtained from experiments [8,15]. The analysis also reported systematic discrepancies between DMI values obtained from bubble expansion and those derived from BLS on the same samples. From an industrial perspective, magneto optical methods such as magneto optical Kerr effect (MOKE) microscopy are particularly attractive because they enable rapid, large-scale analysis compared with more complex and time intensive spectroscopic techniques, like BLS. Kerr microscopy, however, measures only a projection of the magnetization, typically one component at a time, and its spatial resolution is usually on the order of hundreds of nanometers. In principle, DMI information should be encoded in the morphology of bubble domain walls [18–20], but the limited resolution prevents direct access to the internal chiral wall structure. As a result, standard analytical approaches infer DMI only indirectly, most often through domain wall velocities within fitting models. A method capable of extracting DMI from a single, low resolution image would therefore mitigate these constraints and make bubble based DMI characterization more reliable and scalable.

In this setting, data driven inference provides a complementary strategy [5,21]. Rather than isolating a single ideal observable, ML can learn a direct mapping from MOKE images to an effective DMI parameter while accounting for variability in both the sample and the measurement process. Several studies have demonstrated the potential of convolutional neural networks (CNNs) to extract magnetic material parameters from spin texture data [5,21,22]. A common route is to train models on spin configuration images generated by micromagnetic simulations and then transfer the learned representation to experimental observations. While this workflow enables large, systematically controlled training datasets, it also highlights a central challenge: ensuring that models trained on simulated textures generalize reliably to experimentally acquired images. Closing the gap between simulations and reality is essential for using ML as a robust, scalable tool for quantitative magnetic characterization in defect-limited thin films.

In this manuscript, we present a data-driven framework that uses CNNs to estimate DMI strength from synthetic data designed to emulate the quality and spatial resolution of experimental MOKE images. Fig.1 sketches the approach that we have utilized to build this data-driven framework. First, we generate a large dataset from micromagnetic simulations spanning a broad range of DMI values, in-plane magnetic fields, and defect landscapes. We choose the simulated textures to reflect the bubble-like configurations that are typically accessible in MOKE measurements. We introduce experimentally relevant disorder via spatial variations in magnetic anisotropy, which we construct using Voronoi tessellations. We incorporate finite optical resolution by pixelating the simulated magnetization and applying spatial averaging. This dataset allows us to systematically assess DMI inference under increasingly realistic conditions, including measurement noise and reduced spatial resolution. It also motivates a compact CNN architecture that matches the simple, well-defined topology and morphology of bubbles while limiting overfitting. Using this framework, we quantify estimation accuracy as a function of the selected magnetization component, noise level, and degree of pixelation. Our results support ongoing efforts to use ML to quantitatively extract magnetic material parameters from realistic and imperfect experimental observations.

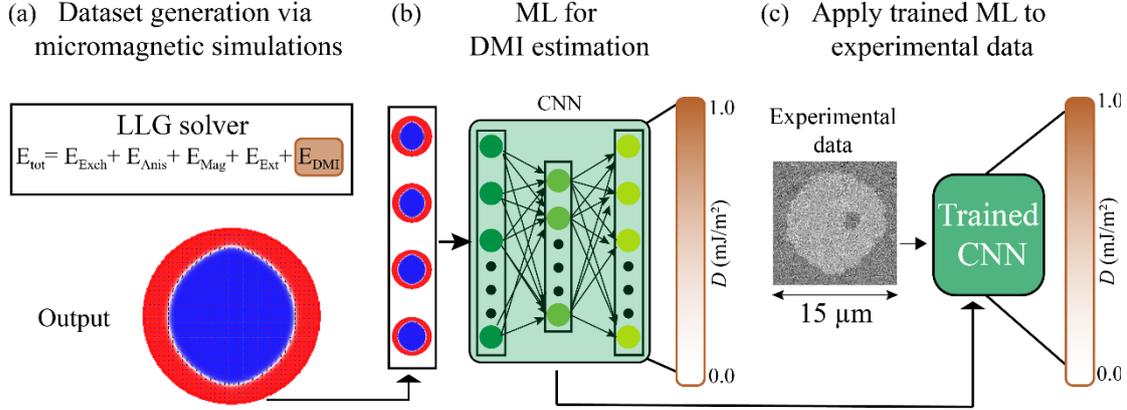

Fig.1. Data-driven framework. (a) Generation of a large dataset via micromagnetic simulations; (b) training of ML approach to extract DMI from simulation data; (c) future perspective: application of the trained model to infer DMI from a single experimental magneto optical image.

**METHODS**

DATASET GENERATION

The first step consists in generating the dataset required to train the CNN. To emulate experimental data obtained through MOKE measurements, we generated bubble-like magnetic textures using micromagnetic simulations, as their profiles are strongly influenced by the DMI strength [8,15,17], see the MOKE image in Fig. 1(c). We considered a circular ferromagnetic film with dimensions 795 nm × 795 nm × 12 nm, using magnetic parameters commonly reported for bubble-like systems [23]: uniaxial magnetic anisotropy density $K_u = 0.33$ MJ/m$^3$ with easy axis along the out-of-plane direction ($\mathbf{e}_z$), saturation magnetization $M_s = 0.55$ MA/m, and exchange stiffness $A_{\text{ex}} = 10$ pJ/m. Simulations are performed at zero temperature.

To investigate the influence of the interfacial DMI, the DMI strength parameter $D$ was varied in the range 0.0 to 1.0 mJ/m$^2$, covering configurations from non-chiral magnetic bubbles to skyrmion-like textures. In order to reproduce realistic experimental conditions and build a generalizable model, structural disorder was introduced through spatial variations of $K_u$, representing pinning sites, surface roughness, and grain boundaries. These inhomogeneities were modeled using Voronoi tessellations to generate spatially correlated anisotropy distributions.

The final equilibrium magnetic configurations were obtained using the in-house GPU-accelerated micromagnetic solver PETASPIN, which numerically integrates the Landau–Lifshitz–Gilbert (LLG) equation of motion [24–27]. The simulation domain was discretized using cells of size 1.5 nm × 1.5 nm × 12 nm. This discretization was selected to ensure computational efficiency while maintaining sufficient spatial resolution to accurately capture the relevant micromagnetic features [23].

The dataset was generated following a systematic protocol. (*i*) The system was initialized in a uniformly saturated state with out-of-plane magnetization along $+m_z = +M_s$. An external magnetic field $H_z$ of opposite polarity was then gradually applied until the nucleation of a magnetic bubble at $H_z = -6$ mT, following a procedure similar to that reported by Liu *et al.* [23]. In the absence of pinning sites, the magnetic configuration would relax back to the uniform state once the field is removed; therefore, a minimum field of $H_z = -6$ mT was maintained to stabilize the bubble. (*ii*) An in-plane magnetic field $H_x = 100$ mT was applied to promote anisotropic expansion of the bubble along the field direction. In the presence of DMI, this expansion becomes asymmetric, providing a measurable signature of the interaction strength. To ensure rapid convergence, we performed a finite size analysis and selected the smallest simulation domain for which the bubble profile was not perturbed by boundary driven effects. (*iii*) The perpendicular magnetic field $H_z$ was varied over a broad range to control the bubble expansion. This was performed incrementally by increasing the magnitude of $H_z$ from $-6$ mT to $-10$ mT in steps of $\Delta H_z = -0.01$ mT. The entire protocol was repeated for each selected value of the DMI strength. A total of 842 images were generated.

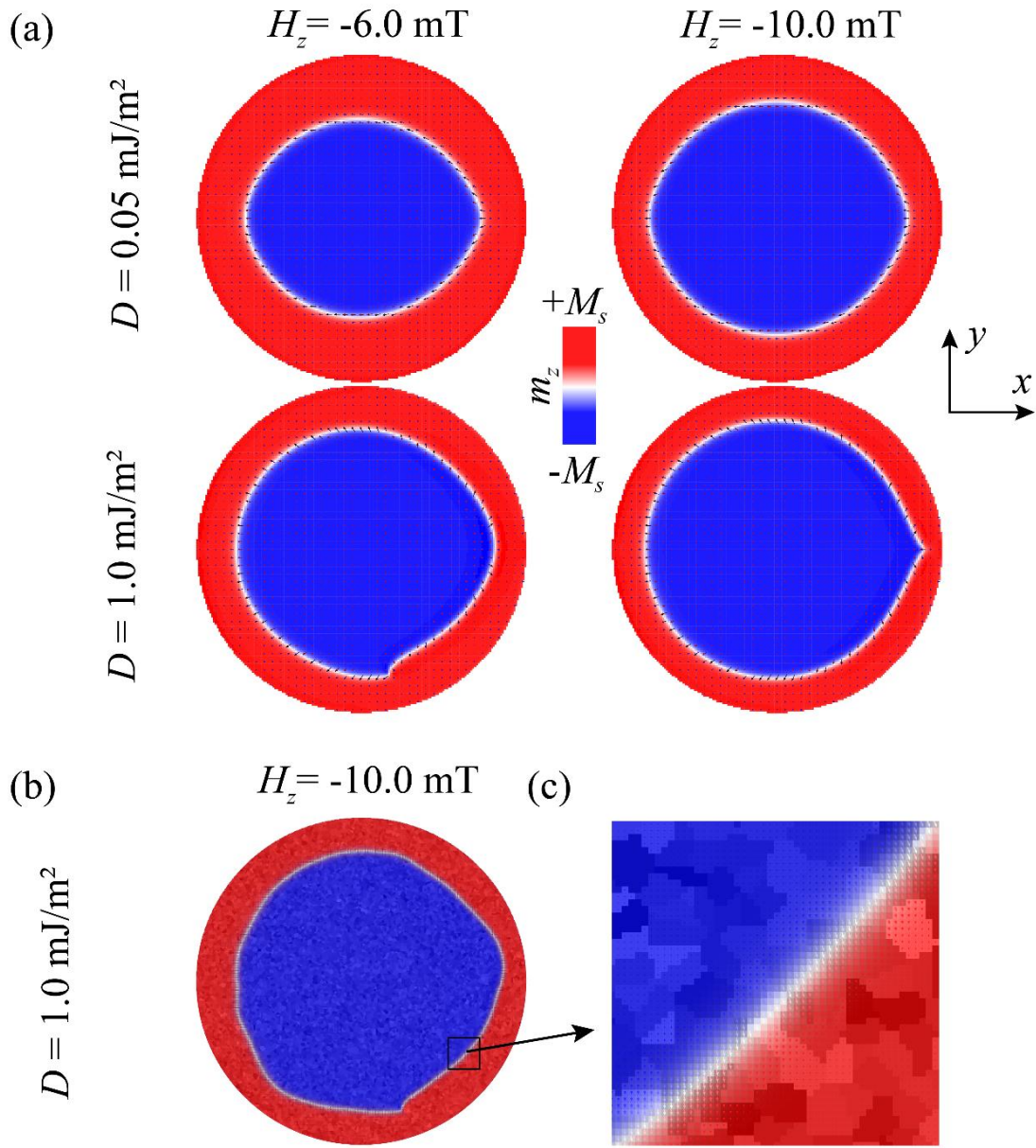

Fig.2. Micromagnetic dataset. (a) Representative magnetization snapshots illustrating the influence of the perpendicular magnetic field $H_z$ and DMI strength $D$ on the magnetic texture. For all figures, we considered $H_x = -100$ mT. The left column corresponds to $H_z = -6$ mT, and the right column to $H_z = -10$ mT. The top row shows $D = 0.05$ mJ/m$^2$, and the bottom row $D = 1.00$ mJ/m$^2$. (b) Example of a magnetic texture in the presence of structural disorder modeled using a Voronoi tessellation. (c) Enlarged view of the domain-wall region highlighted in (b), with the corresponding Voronoi tessellation overlaid for $H_z = -6$ mT and $D = 1.0$ mJ/m$^2$. In all panels, the color map represents the out-of-plane magnetization component $m_z$ (red corresponding to $m_z = +M_s$, blue to $m_z = -M_s$, and white to $m_z = 0$) and arrows indicate the in-plane magnetization direction in each cell.

Fig. 2(a) illustrates the dataset generation protocol. The left column shows representative magnetization snapshots at step (*ii*) for $D = 0.05$ mJ/m$^2$ (top row) and $D = 1.0$ mJ/m$^2$ (bottom row). In both cases, the magnetic bubble elongates along the direction of the applied in-plane field (x axis) due to the interplay between DMI induced chirality and the field driven alignment of the domain wall magnetization along the boundary [28]. At larger DMI, the system can overcome the exchange energy cost to nucleate Bloch lines and domain wall skyrmions, which manifest as pronounced kinks and localized distortions along the bubble boundary [29,30]. The right column displays the magnetization configurations at the final stage of step (*iii*), when the bubble has expanded under $H_z = -10$ mT. In the presence of an in-plane field $H_x$, the expansion becomes asymmetric, with the degree of asymmetry increasing with DMI strength. These DMI-dependent modifications of the bubble profile provide distinctive morphological features that serve as input to the machine-learning analysis.

To incorporate realistic experimental conditions, including pinning, surface roughness, and grain boundaries, we introduced non-uniformity in the uniaxial anisotropy $K_u$ using Voronoi tessellations. The same generation protocol described for the uniform case was then applied to these non-uniform samples. Six independent Voronoi tessellations were generated, with $K_u$ drawn from a normal distribution centered at $K_{u0} = 0.33$ MJ/m$^3$ and a normalized standard deviation of $\sigma_{K_u} = 5\%$ for each grain. A total of 570 images were generated using Voronoi tessellations. Fig. 2(b) shows a representative magnetic configuration obtained for $H_z = -6$ mT and $D = 1.0$ mJ/m$^2$. Compared to the uniform case shown in Fig. 2(a), spatial variations in $K_u$ introduce visible distortions in the bubble profile, reflected in local changes in the out-of-plane magnetization contrast. To further illustrate this effect, Fig. 2(c) presents a magnified view of the domain-wall region highlighted in Fig. 2(b), with the corresponding Voronoi tessellation overlaid. These disorder-induced features modify the magnetic texture, introduce additional structural complexity, and are essential for assessing the reliability and generalization capability of the ML model.

MACHINE LEARNING IMPLEMENTATION

The morphology of the observed magnetic bubble-like textures are determined by the different energy contributions, which can be effectively described by a reduced micromagnetic energy [31,32]. Gradient terms, such as exchange and DMI, depend on the relative orientation of neighboring spins and together determine characteristic length scales, domain wall energy, and chirality. Local terms, such as magnetic anisotropy and external magnetic fields, capture coupling to symmetry breaking fields and external perturbations. They strongly influence the area and shape of the bubble-like configuration. Long range dipolar interactions in case of large domains can be captured by an effective local contribution, for example via a renormalized anisotropy [31,32].

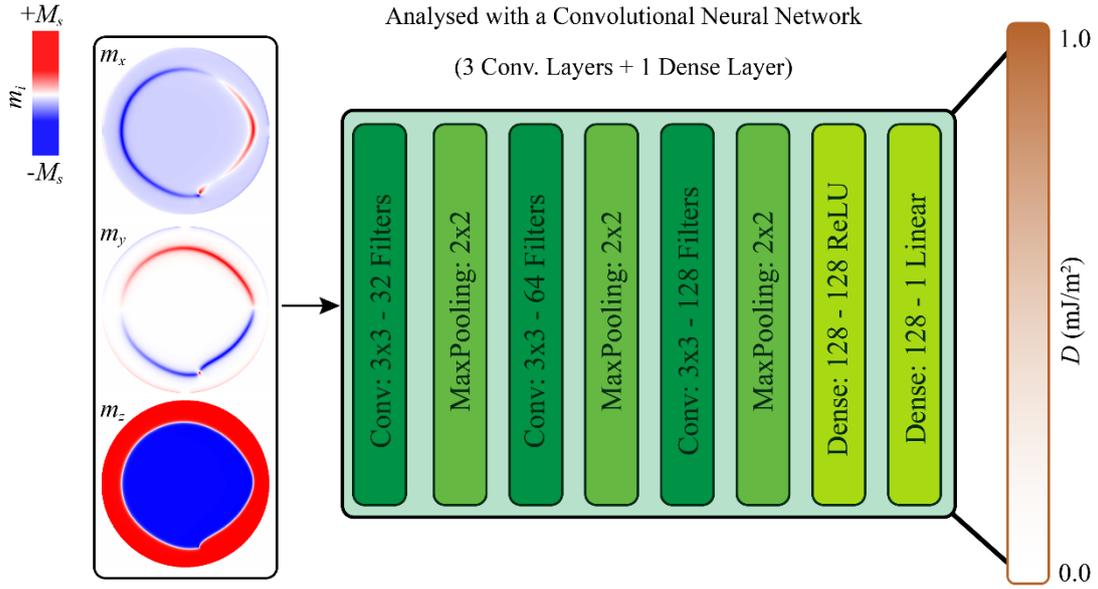

Fig. 3. Machine Learning Convolutional Neuronal Network (CNN) workflow for DMI estimation. Micromagnetic simulation data is stored as text files and used as inputs to the network. The CNN consists of three convolutional blocks, each composed of convolutional layers followed by max-pooling operations, a fully connected dense layer with ReLU activation, and a final linear layer that maps the output to a one-dimensional value corresponding to the predicted DMI strength.

From this phenomenological viewpoint, DMI influences bubble textures through robust, mesoscopic signatures such as the characteristic size, the wall structure, and curvature-related features along the boundary, whereas disorder mainly induces non universal roughness and pinning driven distortions [20,29,31,32]. The designed neural network aims to learn features that correlate with these stable geometric and length scale cues while remaining insensitive to fine scale discrepancies introduced by defects, finite resolution, and noise. For this reason, we adopt a compact CNN architecture that balances expressivity and generalization [33]. Convolutions provide an appropriate inductive bias for texture analysis through local receptive fields and translation equivariance, while limiting the network depth and parameter count reduces the risk of overfitting to disorder realizations or simulation specific artifacts. The latter is particularly important because disorder increases the apparent image complexity without increasing the intrinsic dimensionality of the underlying physics. Therefore, a larger model could achieve low training error by memorizing boundary roughness rather than learning reproducible DMI dependent signatures.

Fig. 3 shows the ML workflow. The inputs to the network are bubble domain-like textures generated by micromagnetic simulations and stored as text files containing the simulated magnetic configuration, with each sample labeled by the DMI value used to produce it. We consider two families of data: a "uniform" dataset that corresponds to simulations with a homogeneous energy landscape, and a "non-uniform" dataset that includes Voronoi textured disorder that mimics spatial variations of material properties. For training and evaluation, the data is randomly shuffled and split into 80% training and 20% testing sets, allowing performance to be assessed on held out textures with defect realizations and bubble morphologies not seen during optimization.

The regression model is implemented in TensorFlow Keras and follows a compact CNN design. It consists of three convolutional blocks with 3×3 kernels and rectified linear unit (ReLU) activations, using 32, 64, and 128 filters, respectively. Each convolutional block is followed by 2×2 max pooling, which progressively reduces spatial resolution while increasing the effective receptive field, so that the network can integrate information along the bubble boundary at multiple length scales. After the convolutional stack, we apply global average pooling to compress the feature maps into a compact latent representation. The pooled representation is then passed to a fully connected layer with 64 neurons and ReLU activation, and finally to a single linear output neuron that returns the predicted DMI value.

Training is performed with the Adam optimizer by minimizing the mean squared error between predicted and true DMI values. In addition to the training loss, we track the mean absolute error (MAE) as a more directly interpretable measure of regression accuracy. Model performance is summarized using both the MAE and the standard deviation of the prediction error computed separately on the training and test sets. For each analysis, we trained from 10 randomly initialized weights and trained for 30 epochs.

**RESULTS**

To validate the proposed CNN framework for estimating the interfacial DMI strength from magnetic bubble images, we performed a systematic analysis addressing robustness, generalization, spatial resolution, and noise sensitivity.

We first evaluated the sensitivity of the CNN to sample uniformity. Three datasets were considered: uniform samples, non-uniform samples with grain-induced anisotropy variations, and the combined dataset. For the uniform samples, DMI values were varied in steps of 0.01 mJ/m², while for the non-uniform samples a coarser variation of 0.1 mJ/m² was used. Fig. 4 summarizes the MAE for each configuration, including the magnetization components used as input and the degree of sample uniformity. In this manuscript, the units of the MAE and the associated standard deviation (std) are given in mJ/m². For each dataset, the network was trained using 10 independent random initializations over 30 epochs to ensure statistical reliability.

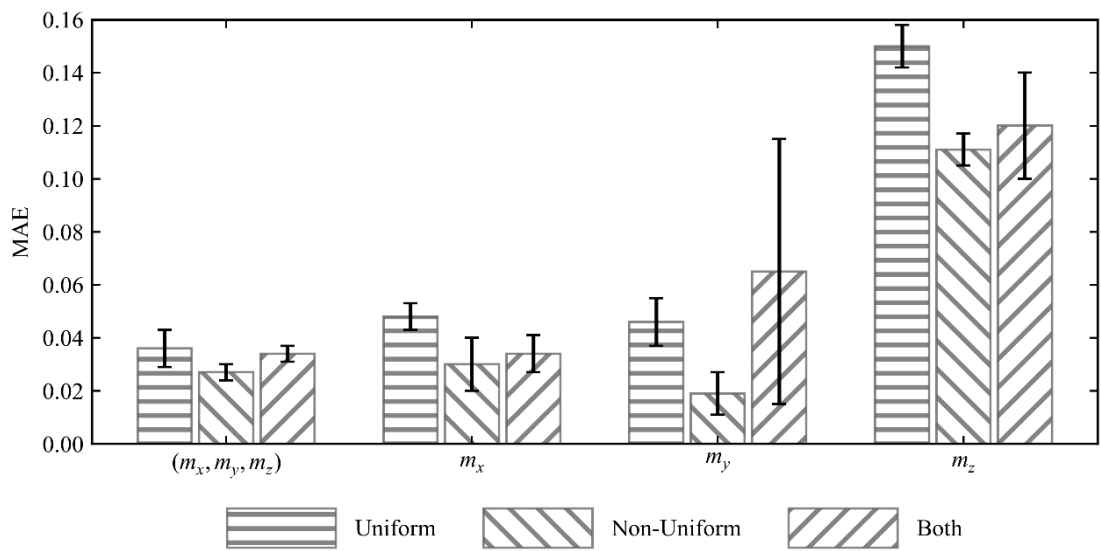

Fig. 4. Mean absolute error (MAE) as a function of sample uniformity.

The results indicate that the in-plane magnetization components yield the lowest MAE, whereas the out-of-plane component $m_z$ produces the largest error. This behavior is consistent with the physical role of DMI, which primarily affects the in-plane rotation of magnetization at the skyrmion boundary. Consequently, information encoded in $m_z$ alone does not fully capture the chiral boundary structure induced by DMI. Combining all three magnetization components further improves accuracy. Interestingly, training exclusively on non-uniform samples leads to improved performance, which can be attributed to the enhanced sensitivity of boundary curvature and grain-induced distortions to variations in DMI strength.

An essential property of neural networks is their ability to generalize beyond the training distribution. Since the bubble domain profile varies smoothly with DMI, we performed a generalization test by partitioning the DMI range into disjoint training and test intervals. Specifically, we considered three scenarios, presented in Fig. 5(a): (*i*) training in the range (0.05, 0.6) and testing in (0.6, 1.0); (*ii*) training in (0.4, 1.0) and testing in (0.05, 0.4); and (*iii*) training in (0.2, 0.8) and testing in the complementary intervals (0.05, 0.2) and (0.8, 1.0). In each case, 100% of the images within the training interval, which is about two-thirds of the dataset, were used for training and the remaining 100% for testing. The MAE values reported in Fig. 5(b) demonstrate robust generalization across all partitions, with weak dependence on the selected DMI interval. This robustness reflects the smooth and monotonic dependence of the bubble profile on the DMI strength.

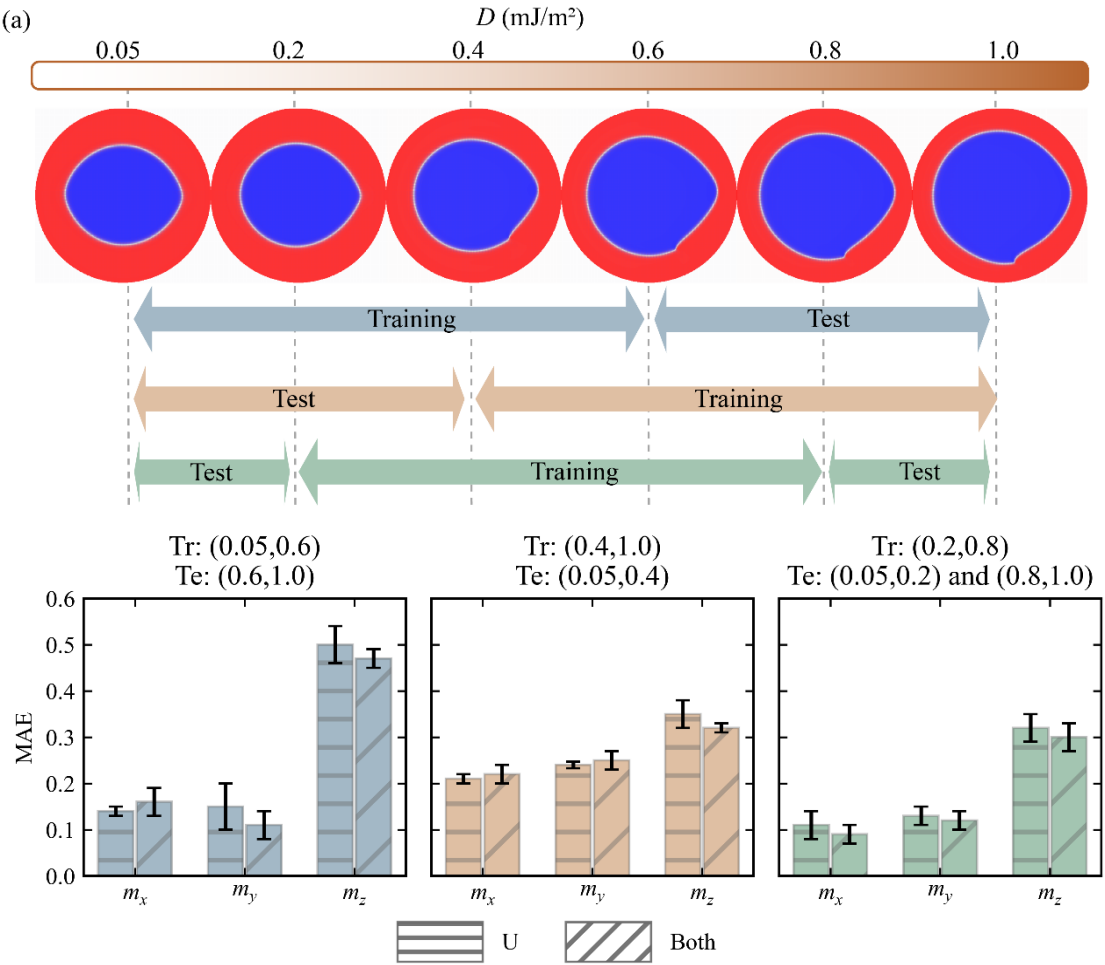

Fig. 5. Mean absolute error (MAE) as a function of training and test ranging. (a) Schematic representation of the DMI intervals used for training and testing, together with the corresponding

evolution of the bubble configuration as a function of DMI strength. (b) Histograms of the MAE for different magnetization components and for the selected training and test ranges. "Tr" denotes the training set and "Te" the test set. "U" denoted only the dataset with uniform background and "Both" means the full dataset, including uniform and non-uniform backgrounds.

We next evaluated the influence of spatial resolution by systematically reducing image resolution through pixelation. In preprocessing, each magnetization image was coarse grained by block averaging. The magnetization components were grouped into non overlapping $b \times b$ pixel blocks, and each block was replaced by the area averaged magnetization vector. A schematic illustration of the process is shown in Fig. 6(a). This reduces the effective image size to $(H/b) \times (W/b)$ and mimics the loss of spatial detail expected from limited experimental resolution. We performed this analysis for pixelation factors $b = 2, 5, 10$, and $20$. Fig. 6(b) reports the corresponding MAE values for the full dataset and for different magnetization components. The CNN remains highly robust to resolution degradation up to a pixelation factor of 10. For a factor of 20, a slight improvement in accuracy is observed. This behavior suggests that the DMI signature is encoded predominantly in global structural features rather than in fine-scale local variations. Therefore, these results indicate that the method is well suited for experimental data with limited spatial resolution.

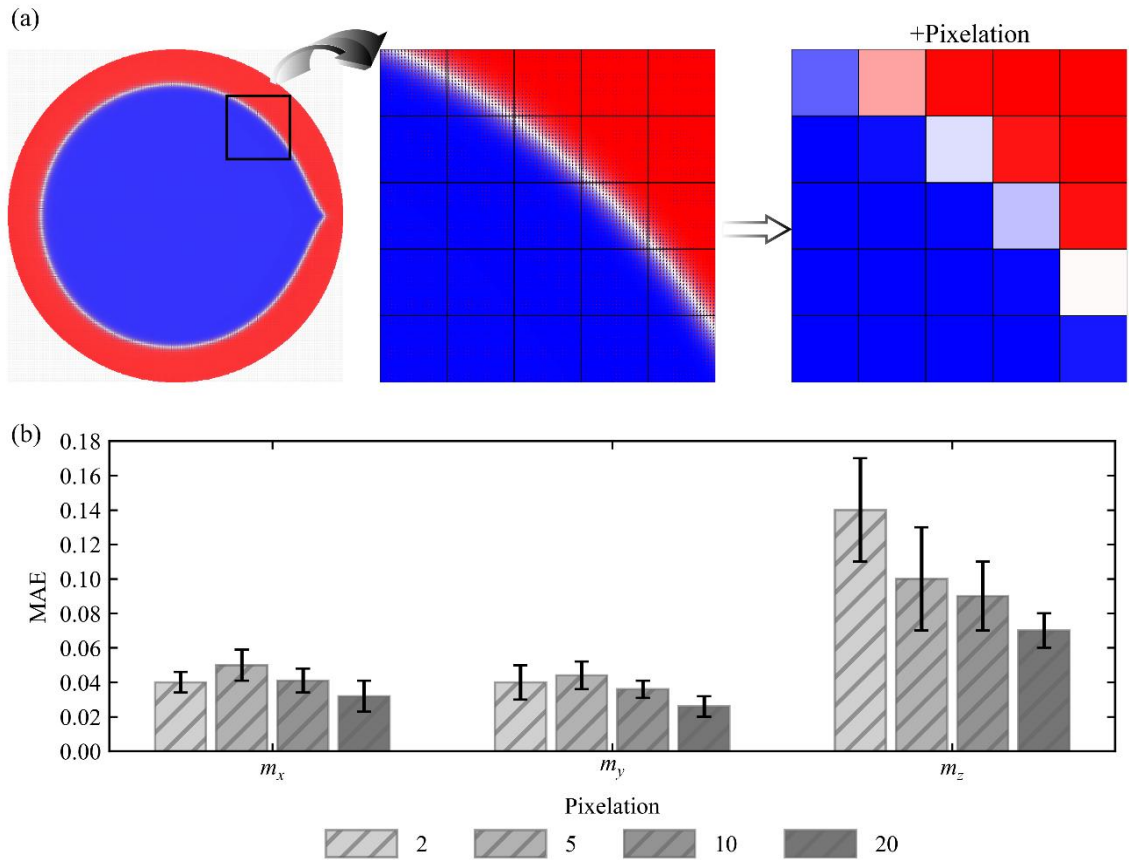

Fig. 6. Mean absolute error (MAE) as a function of pixelation. (a) Schematic illustration of the pixelation process, showing a magnified region of the analyzed image and the corresponding effect of pixelation on that region. (b) Histograms of the MAE as a function of the magnetization component and the pixelation factor, evaluated over the full dataset.

Finally, we assessed robustness against additive noise. The dataset was augmented fivefold by adding Gaussian noise with fixed standard deviation $\sigma$, followed by analysis using pixelation factors of 2 and 5. A schematic illustration of the process is shown in Fig. 7(a). Fig. 7(b) summarizes the MAE for $\sigma = 0.05$, $0.10$, and $0.15$. Within the statistical uncertainty of the training procedure, no significant degradation in accuracy was observed. These results demonstrate that the CNN-based approach maintains stable performance under realistic noise levels and further support its applicability to experimental measurements.

**CONCLUSIONS**

We have proposed and validated a machine-learning-based framework for estimating the interfacial DMI strength directly from magnetic bubble profiles. A compact CNN was trained using datasets generated through micromagnetic simulations and systematically evaluated under progressively realistic conditions. Specifically, we analyzed the influence of structural non-uniformity, pixelation, and additive noise, as well as the generalization capability of the network across disjoint DMI ranges.

Our results show that the in-plane magnetization components provide the highest predictive accuracy, consistent with the fact that DMI primarily governs the chiral rotation of magnetization at the bubble boundary. The method remains robust under reduced spatial resolution and moderate noise levels and exhibits reliable generalization beyond the training interval. These findings demonstrate that the proposed approach can operate under realistic experimental constraints and support the integration of ML techniques into experimental workflows to enhance the quantitative determination of DMI parameters. Further work will explore the application of the approach to experimental data.

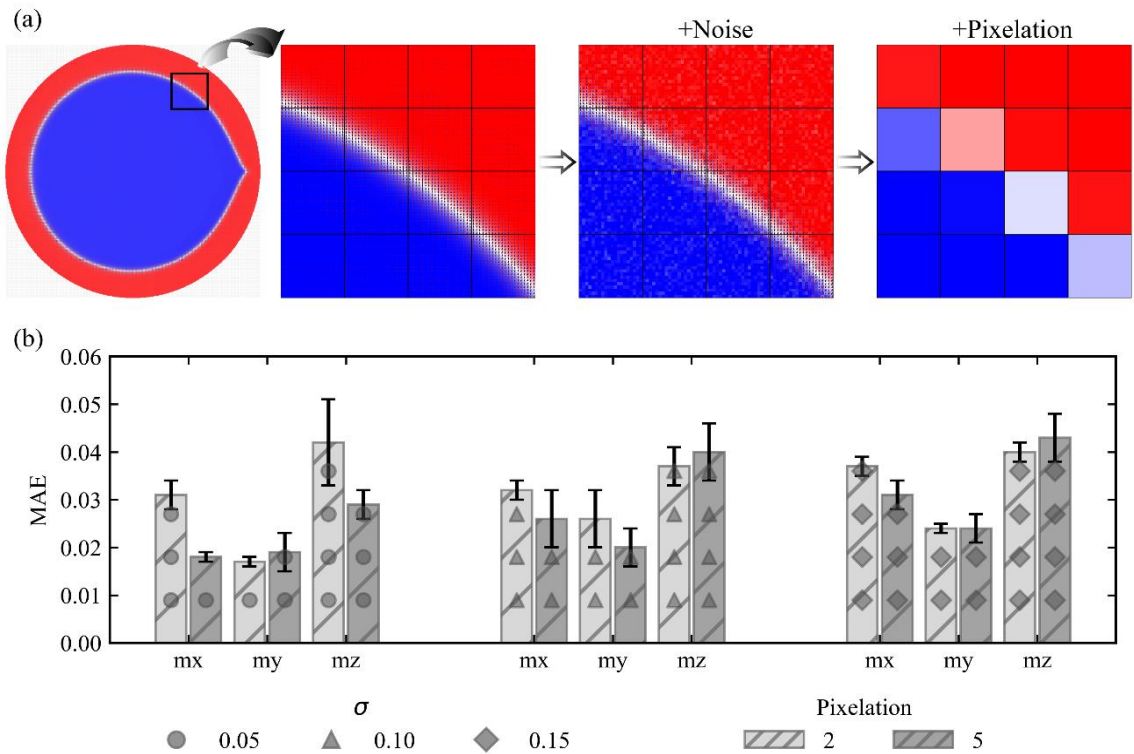

Fig. 7. Mean absolute error (MAE) as a function of added noise. (a) Schematic illustration of the noise addition and pixelation procedure, showing a magnified region of the analyzed image and the corresponding effect of additive noise and pixelation on that region. (b) MAE as a function of the magnetization component, the noise level characterized by the standard deviation $\sigma$, and the pixelation factor, evaluated over the full dataset.

## ACKNOWLEDGEMENTS
The work was supported by the MetroSpin project, funded by PRIN2022 SAYARY (CUP E53D2300183 0006) financed by the European Union - Next Generation EU and MUR (Ministero dell'Universita e della Ricerca). D.R, A.H., and G.F. acknowledge participation in TOPOCOM, which is funded by the European Union's Horizon Europe Programme Horizon 1.2 under the Marie Sklodowska-Curie Actions (MSCA), Grant Agreement No. 101119608. D.R, A.M., A.H, E.P., G.F., M.C., V.P are with the PETASPIN team and thank the PETASPIN association (www.petaspin.com).